\documentclass{article}
\usepackage{graphicx}  
\title{A de Broglie-Bohm Like Model for Dirac Equation\\submited to Nuovo Cimento}
\author{O. Chavoya-Aceves\\
Camelback H. S., Phoenix, Arizona, USA.\\
chavoyao@yahoo.com}

\begin{document}

\maketitle

\begin{abstract}
A de Broglie-Bohm like model of Dirac equation, that leads to the
correct Pauli equations for electrons and positrons in the
low-speed limit, is presented. Under this theoretical framework,
that affords an interpretation of the \emph{quantum potential},
the main assumption of the de Broglie-Bohm theory---that the local
momentum of particles is given by the gradient of the phase of the
wave function---wont be accurate. Also, the number of particles
wont be locally conserved. Furthermore, the representation of
physical systems through wave functions wont be complete.

\end{abstract}

\section{Introduction}
We'll show that Dirac equation can be interpreted as describing
the motion of four currents of particles, under the only action of
the electromagnetic field, with four-velocities given by the
equations
\[
m v_{i,\mu}=-\partial_\mu\Phi_i-\partial_\mu S_i \pm
\frac{q}{c}A_\mu,
\]
where $\pm S_i$ are the phases of the components of the wave
function, in such way that there are not negative energies, and
the functions $\Phi_i$ are \emph{hidden variables}. This leads to
the correct Pauli equations for electrons and positrons in the
low-speed limit, whenever the \emph{zitterbegwegung} terms can be
neglected.

Within the theoretical framework of this work, the main assumption
of the de Broglie-Bohm theory of quantum mechanics \cite{BOHM} is
not valid and a picture of particles moving under the action of
electromagnetic field alone, without any quantum potential,
emerges. However, the number of particles is not locally
conserved. Which is very well known, as a matter of fact. Also,
given that \emph{hidden variables} have been introduced---on the
grounds of some general electrodynamic considerations, given in
the first section---the representation of physical systems through
wave functions wont be complete and, therefore, as foreseen by
Einstein, Podolsky, and Rosen, quantum mechanics wont be a
complete theory of motion.

\section{On the Motion of Particles Under the Action of Electromagnetic Field}
Consider a non-stochastic ensemble of particles whose motion is
described by means of a function
\[
\vec{r}=\vec{r}(\vec{x},t),
\]
such that $\vec{r}(\vec{x},t)$ represents the position, at time
$t$, of the particle that passes through the point $\vec{x}$ at
time \emph{zero}---the $\vec{x}$-particle. This representation of
motion coincides with the lagrangian representation used in
hydrodynamics\cite{ARIS}.

As to the function $\vec{r}(\vec{x},t)$, we suppose that it is
invertible for any value of $t$. In other words, that there is a
function $\vec{x}=\vec{x}(\vec{r},t)$, that gives us the
coordinates, at time zero, of the particle that passes through the
point $\vec{r}$ at time $t$. Also, we suppose that
$\vec{r}(\vec{x},t)$ is a continuous function, altogether with its
derivatives of as higher order as needed to secure the validity of
our conclusions.

According to the definition of $\vec{r}(\vec{x},t)$, the velocity
of the $\vec{x}$-particle at time $t$ is
\[
\vec{u}(\vec{x},t)=\left( \frac{\partial \vec{r}}{\partial
t}\right)_{\vec{x}}.
\]

Writing $\vec{x}$ as a function of $\vec{r}$ and $t$, we get the
velocity field at time $t$:
\[
\vec{v}(\vec{r},t)=\vec{u}(\vec{x}(\vec{r},t),t).
\]

Lets consider the corresponding four-velocity:
\[
v^\mu=\left(\frac{c}{\sqrt{1-v^2/c^2}},\frac{\vec{v}}{\sqrt{1-v^2/c^2}}\right)
\]
Using four-dimensional tensorial notation, the derivative of
$v_\mu$ with respect to the proper time, along the world-line of
the corresponding particle is:
\[
\frac{d v_\mu}{ds}=v^\nu \partial_\nu v_\mu
\]
From the identity
\begin{equation}\label{square of four-velocity}
  v^\nu v_\nu=c^2
\end{equation}
we prove that
\[
v^\nu\partial_\mu v_\nu=0,
\]
that allows us to write
\begin{equation}\label{four acceleration}
  \frac{d v_\mu}{ds}=(\partial_\nu v_\mu -\partial_\mu
  v_\nu)v_\nu.
\end{equation}

If the four-force acting on the ensemble of particles has an
electromagnetic origin, it is given by the expression
\begin{equation}\label{lorentz}
  f_\mu=\frac{q}{c}F_{\mu\nu}v^\nu,
\end{equation}
where
\begin{equation}\label{faraday}
  F_{\mu\nu}=\partial_\mu A_\nu-\partial_\nu A_\mu
\end{equation}
is Faraday's tensor, and $A_\mu$ is the electrodynamic
four-potential.

From the law of motion
\[
m\frac{dv_\mu}{ds}=f_\mu,
\]
and equations (\ref{four acceleration}) to (\ref{faraday}) we
prove that
\begin{equation}\label{fundamental equation}
  (\partial_\mu p_\nu - \partial_\nu p_\mu)v^\nu=0,
\end{equation}
where
\begin{equation}\label{four-momentum}
  p_\mu=m v_\mu + \frac{q}{c}A_\mu.
\end{equation}

There is a class of solutions of equations (\ref{fundamental
equation}) where $p_\mu$ is the four-gradient of a function of
space-time coordinates
\begin{equation}\label{definition of mechanical action}
  p_\mu=-\partial_\mu \phi.
\end{equation}
Therefore,
\[
\partial_\mu p_\nu -\partial_\nu p_\mu =0,
\]
and (\ref{fundamental equation}) is obviously satisfied.

Equation (\ref{four-momentum}) can be written in the form
\begin{equation}\label{kinetic momentum}
  m v_\mu=-\partial_\mu \phi -\frac{q}{c} A_\mu
\end{equation}

If the four-potential meets the Lorentz condition,
\begin{equation}\label{lorentz condition}
  \partial^\mu A_\mu =0,
\end{equation}
equation (\ref{kinetic momentum}) is analogous to the
decomposition of a classical, three-dimensional field, into an
irrotational and a solenoidal part. The electromagnetic field
appears thus as determining the four-dimensional vorticity of the
field of \emph{kinetic momentum} $m v_\mu$.

Equation (\ref{square of four-velocity}) tells us that functions
$\phi$ are not arbitrary, but are subject to the condition:
\begin{equation}\label{relativistic hamilton-jacobi}
  (\partial^\mu \phi
  +\frac{q}{c}A^\mu)(\partial_\mu\phi+\frac{q}{c}A_\mu)=m^2 c^2,
\end{equation}
which is the relativistic Hamilton-Jacobi equation.\cite[Ch.
VIII]{LANCZOS}

The components of the kinetic momentum are:
\begin{equation}\label{components of kinetic momentum}
  \left(\frac{K}{c},\vec{p}\right)=\left(-\frac{1}{c}\frac{\partial \phi}{\partial t}-\frac{q}{c}V,\nabla \phi -
  \frac{q}{c}\vec{A}\right),
\end{equation}
where $V$ and $\vec{A}$ are the components of the electrodynamic
four-potential.

In the low-speed limit, we have
\[
K\approx mc^2 + \frac{p^2}{2m}.
\]
From this and (\ref{components of kinetic momentum}) we get:
\begin{equation}\label{non-relativistic hamilton-jacobi}
  \frac{\partial \phi}{\partial t}+ \frac{(\nabla \phi -
  \frac{q}{c}\vec{A})^2}{2m}+ q V + m c^2 =0,
\end{equation}
that, but for the constant term $m c^2$, is the non-relativistic
Hamilton-Jacobi equation.

\section{A de Broglie-Bohm Like Interpretation of Dirac Equation}
Lets consider Dirac equation for particles of mass $m$ and charge
$q$, under the action of an electromagnetic field that meets the
Lorentz condition(\ref{lorentz condition}):
\begin{equation}\label{dirac equation}
  \gamma^\mu (i\hbar \partial_\mu -\frac{q}{c} A_\mu)\psi=mc\psi
\end{equation}
or
\begin{equation}\label{dirac equation adjoint}
\overline{\psi}\gamma^\mu(-i\hbar\overleftarrow{\partial}_\mu-\frac{q}{c}A_\mu)=mc\overline{\psi}
\end{equation}
where $\overline{\psi}=\psi^\dag\gamma^0$.

It is easy to show that:
\begin{equation}\label{continuity dirac}
  \partial_\mu \iota^\mu=0
\end{equation}
where
\begin{equation}\label{dirac current}
  \iota^\mu=c\overline{\psi}\gamma^\mu\psi.
\end{equation}
From (\ref{dirac equation} \& \ref{dirac equation adjoint})
\[
\overline{\psi}\gamma^\mu\gamma^\nu(i\hbar\partial_\nu-\frac{q}{c}A_\nu)\psi=mc\overline{\psi}\gamma^\mu\psi,
\]
\[
\overline{\psi}\gamma^\nu\gamma^\mu(-i\hbar\overleftarrow{\partial}_\nu-\frac{q}{c}A_\nu)\psi=mc\overline{\psi}\gamma^\mu\psi.
\]
Therefore:
\begin{equation}\label{total current one}
  \iota^\mu=\frac{i\hbar(\overline{\psi}\gamma^\mu\gamma^\nu\partial_\nu\psi-\partial_\nu\overline{\psi}\gamma^\nu\gamma^\mu
  \psi)}{2m}-\frac{q}{mc}\overline{\psi}\psi A^\mu
\end{equation}

Considering that:
\begin{equation}\label{anticommutation}
  \{\gamma^\mu,\gamma^\nu\}=2g^{\mu\nu}\textbf{I},
\end{equation}
we can prove now that
\begin{equation}\label{total current two}
   \iota^\mu=\frac{i\hbar(\overline{\psi}\partial^\mu\psi-\partial^\mu\overline{\psi}\psi)}{2m}+\frac{ih\partial_\nu(\overline{\psi}[\gamma^\mu,\gamma^\nu]\psi)}{4m}-\frac{q}{mc}\overline{\psi}\psi A^\mu
\end{equation}
The divergence of the second term of the right side is zero.
Therefore we guess that the current of particles could be given
by:
\begin{equation}\label{electric current one}
j_0^\mu=\frac{i\hbar(\overline{\psi}\partial^\mu\psi-\partial^\mu\overline{\psi}\psi)}{2m}-\frac{q}{mc}\overline{\psi}\psi
A^\mu
\end{equation}
which is made of four terms:
\begin{equation}\label{current decomposition}
  mj^\mu_{0,1}=\rho_1(-\partial^\mu S_1-\frac{q}{c}A^\mu)
\end{equation}
\[
  mj^\mu_{0,2}=\rho_2(-\partial^\mu S_2-\frac{q}{c}A^\mu)
\]
\[
  mj^\mu_{0,3}=\rho_3(-\partial^\mu S_3+\frac{q}{c}A^\mu)
\]
\[
  mj^\mu_{0,4}=\rho_4(-\partial^\mu S_4+\frac{q}{c}A^\mu)
\]
where
\begin{equation}\label{madelung transformation}
  \psi_1=\sqrt\rho_1e^{\frac{i}{\hbar}S_1}
\end{equation}
\[
 \psi_2=\sqrt\rho_2 e^{\frac{i}{\hbar}S_2}
\]
\[
\psi_3=\sqrt\rho_3 e^{-\frac{i}{\hbar}S_3}
\]
\[
\psi_4=\sqrt\rho_4e^{-\frac{i}{\hbar}S_4}
\]
are the components of the wave function.

The current (\ref{electric current one}) is thus seen as the sum
of four fluxes. According to (\ref{four-momentum}), the first two
can be interpreted as fluxes of particles of charge $q$, while the
last would represent fluxes of particles of charge $-q$. Equations
(\ref{current decomposition}) suggest that we could define the
corresponding four-velocities from:
\begin{equation}\label{wrong forvelocities}
  mw^\mu_1=-\partial^\mu S_1-\frac{q}{c}A^\mu
\end{equation}
\[
  mw^\mu_2=-\partial^\mu S_2-\frac{q}{c}A^\mu
\]
\[
  mw^\mu_3=-\partial^\mu S_3+\frac{q}{c}A^\mu
\]
\[
  mw^\mu_4=-\partial^\mu S_4+\frac{q}{c}A^\mu
\]
but this is not sound, because in such case we should have:
\[
 w^\mu_iw_{i,\mu}=c^2,
\]
which is not true.

However, given that those four-vectors have already the required
four-vorticity---for particles under the action of an
electromagnetic field---, we suppose that there exist functions
$\Phi_1$, $\Phi_2$, $\Phi_3$, and $\Phi_4$, such that the
four-velocities are given by the equations:
\begin{equation}\label{fourvelocities}
  mv^\mu_1=-\partial^\mu (S_1+\Phi_1)-\frac{q}{c}A^\mu,
\end{equation}
\[
 mv^\mu_2=-\partial^\mu (S_2+\Phi_2)-\frac{q}{c}A^\mu,
\]
\[
mv^\mu_3=-\partial^\mu (S_3+\Phi_3)+\frac{q}{c}A^\mu,
\]
\[
mv^\mu_4=-\partial^\mu (S_4+\Phi_4)+\frac{q}{c}A^\mu,
\]
in such way that
\begin{equation}\label{normalization of fourvelocity}
  v^\mu_iv_{i,\mu}=c^2
\end{equation}

The divergence of the current of particles:
\begin{equation}\label{final current}
  j^\mu=\sum_{i=1\cdots 4}\rho_i v^\mu_i
\end{equation}
---that is not proportional to the electrical current---will then be:
\[
\partial^\mu j_\mu =\sum_{i=1,4}\partial^\mu(\rho_i \partial_\mu
\Phi_i)/m,
\]
which is not zero, and, consequently, the number of particles wont
be locally conserved.

From Dirac equation we can show that
\begin{equation}\label{klein gordon}
  -\hbar^2
  \partial^\mu\partial_\mu\psi-\frac{2i\hbar q}{c}A^\mu\partial_\mu\psi+\frac{q^2}{c^2}A^\mu A_\mu\psi=m^2c^2\psi%
  +\frac{i\hbar q}{4c}[\gamma^\mu,\gamma^\nu]F_{\mu\nu}\psi,
\end{equation}
where $F_{\mu\nu}$ is Faraday's tensor or:

\begin{equation}\label{klein gordon first}
  -\hbar^2
  \partial^\mu\partial_\mu\varphi-\frac{2i\hbar q}{c}A^\mu\partial_\mu\varphi+\frac{q^2}{c^2}A^\mu A_\mu\varphi=m^2c^2\varphi%
+\frac{\hbar
q}{c}(\vec{\sigma}\cdot\vec{H}\varphi+i\vec{\sigma}\cdot
\vec{E}\chi ),
\end{equation}

\begin{equation}\label{klein gordon second}
  -\hbar^2
  \partial^\mu\partial_\mu\chi-\frac{2i\hbar q}{c}A^\mu\partial_\mu\chi+\frac{q^2}{c^2}A^\mu A_\mu\chi=m^2c^2\chi%
  +\frac{\hbar q}{c}(\vec{\sigma}\cdot\vec{H}\chi+\vec{\sigma}\cdot \vec{E}\varphi),
\end{equation}
where we have written the wave function in the form:
\begin{equation}\label{analysis of the wave function}
  \psi=\left(%
  \begin{array}{c}
    \varphi \\
    \chi \
  \end{array}
  \right)
\end{equation}

Let's consider a component of $\psi$:
\[
\omega=\sqrt\rho e^{\pm\frac{i}{\hbar}S},
\]

The terms in the left side of equation (\ref{klein gordon first})
are
\[
-\frac{2i\hbar q}{c}A^\mu\partial_\mu\omega=\left(%
\pm\frac{2q\sqrt\rho}{c}A^\mu\partial_\mu S - \frac{2 i \hbar
q}{c}A^\mu\partial_\mu\sqrt\rho \right)e^{\pm\frac{i}{\hbar}S},
\]
and
\[
-\hbar^2\partial^\mu\partial_\mu \omega=
[(-\hbar^2\partial^\mu\partial_\mu \sqrt\rho+\sqrt\rho\partial^\mu
S\partial_\mu S)\mp i\hbar( 2\partial^\mu\sqrt\rho \partial_\mu
S+\sqrt\rho\partial^\mu
\partial_\mu S)]e^{\pm\frac{i}{\hbar}S},
\]

From this we can separate the real and imaginary parts of the
equation that results from left multiplying (\ref{klein gordon
first}) by $\varphi^\dag$:
\begin{equation}\label{real part first}
\sum_{i=1,2} \rho_i\left((-\partial^\mu
S_i-\frac{q}{c}A^\mu)(-\partial_\mu
S_i-\frac{q}{c}A_\mu)-m^2c^2\right)=
\end{equation}
\[
= \sum_{i=1,2}\hbar^2(\sqrt\rho_i\partial^\mu\partial_\mu
\sqrt\rho_i)+\frac{\hbar q}{c}\varphi^\dag
\vec{\sigma}\cdot\vec{H}\varphi+\frac{\hbar q}{c} Re(i\varphi^\dag
\vec{\sigma}\cdot \vec{E}\chi)
\]
and
\begin{equation}\label{imag part first}
\sum_{i=1,2}
\partial^\mu(\rho_i(-\partial_\mu S_i-\frac{q}{c}A_\mu)=\frac{q}{c}Im(i\varphi^\dag\vec{\sigma}\cdot\vec{E}\chi)
\end{equation}

In a similar fashion:
\begin{equation}\label{real part second}
\sum_{i=3,4} \rho_i\left((-\partial^\mu
S_i+\frac{q}{c}A^\mu)(-\partial_\mu
S_i+\frac{q}{c}A_\mu)-m^2c^2\right)=
\end{equation}
\[
= \sum_{i=3,4}\hbar^2(\sqrt\rho_i\partial^\mu\partial_\mu
\sqrt\rho_i)+\frac{\hbar q}{c}\chi^\dag \vec{\sigma}\cdot \vec{H}
\chi + \frac{\hbar q}{c} Re(i\chi^\dag \vec{\sigma}\cdot
\vec{E}\phi)
\]
and
\begin{equation}\label{imag part second}
\sum_{i=3,4}
\partial^\mu(\rho_i(-\partial_\mu S_i+\frac{q}{c}A_\mu)=-\frac{q}{c}Im(i\varphi^\dag\vec{\sigma}\cdot\vec{E}\chi)
\end{equation}

Adding equations (\ref{imag part first} \& \ref{imag part
second}), considering that $\vec{\sigma}\cdot \vec{E}$ is an
hermitian operator, we obtain the equation:
\[
\partial^\mu j_{0,\mu}=0,
\]
which was already known.

From equations (\ref{fourvelocities}):
\begin{equation}\label{key argument}
  (-\partial^\mu S_i \pm \frac{q}{c}A^\mu)(-\partial_\mu S_i \pm
  \frac{q}{c}A_\mu)=(mv^\mu_i+\partial^\mu\Phi_i)(mv_{i,\mu}+\partial_\mu\Phi_i)=
\end{equation}
\[
m^2c^2+2mv_i^\mu\partial_\mu\Phi_i+\partial^\mu\Phi_i\partial_\mu\Phi_i
\]
This allows us rewrite equations (\ref{real part first} \&
\ref{real part second}) in the form:
\begin{equation}\label{real part first final}
\sum_{i=1,2}
\rho_i(v_i^\mu\partial_\mu\Phi_i+\frac{\partial^\mu\Phi_i\partial_\mu\Phi_i}{2m})=%
\sum_{i=1,2}\frac{\hbar^2}{2m}\sqrt\rho_i\partial^\mu\partial_\mu
\sqrt\rho_i+
\end{equation}
\[
\frac{\hbar q}{2mc}\varphi^\dag \vec{\sigma}\cdot \varphi
+\frac{\hbar q}{2mc} Re(i\varphi^\dag \vec{\sigma}\cdot
\vec{E}\chi),
\]
\begin{equation}\label{real part second final}
\sum_{i=3,4}
\rho_i(v_i^\mu\partial_\mu\Phi_i+\frac{\partial^\mu\Phi_i\partial_\mu\Phi_i}{2m})=%
\sum_{i=3,4}\frac{\hbar^2}{2m}\sqrt\rho_i\partial^\mu\partial_\mu
\sqrt\rho_i+
\end{equation}
\[
\frac{\hbar q}{2mc}\chi^\dag \vec{\sigma}\cdot \chi +\frac{\hbar
q}{2mc} Re(i\chi^\dag \vec{\sigma}\cdot \vec{E}\varphi),
\]

In the low-speed limit:
\begin{equation}\label{limit conditions}
  \begin{array}{cc}
    v^0\approx c & (v^1,v^2,v^3)\rightarrow \vec{v} \,
  \end{array}
\end{equation}
where $\vec{v}$ is the classical velocity. Therefore:
\begin{equation}\label{classical limit}
  v^\mu\partial_\mu\Phi+\frac{\partial^\mu\Phi\partial_\mu\Phi}{2m}\approx
\end{equation}
\[
\frac{\partial \Phi}{\partial
t}+\vec{v}\cdot\nabla\Phi-\frac{(\nabla \Phi)^2}{2m}=
\]
\[
\frac{\partial \Phi}{\partial
t}+(\vec{v}-\frac{\nabla\Phi}{m})\cdot \nabla\Phi+\frac{(\nabla
\Phi)^2}{2m}
\]

In this limit also:
\begin{equation}\label{expression for classical velocity}
  \vec{v}=\nabla S + \nabla \Phi \mp \frac{q}{c}\vec{A}
\end{equation}
---where $\vec{A}$ is the vector potential of the electromagnetic
field---from equations (\ref{real part first final}, \ref{real
part second final} \& \ref{classical limit}), we get an
approximation for the kinetic energy:
\begin{equation}\label{energia electron}
K_\varphi=%
\sum_{i=1,2}%
\rho_i( -\frac{\partial \Phi_i}{\partial t}+ mc^2 +\frac{(\nabla
S_i -\frac{q}{c}\vec{A} )^2}{2m}%
-\frac{\hbar^2}{2m}\frac{\Delta\sqrt\rho_i}{\sqrt\rho_i}+\frac{\hbar
q }{2mc}\frac{\varphi^\dag\vec\sigma\cdot\vec
H\varphi}{\varphi^\dag\varphi}+U_\varphi)
\end{equation}
and
\begin{equation}\label{energia positron}
K_\chi=%
\sum_{i=3,4}%
\rho_i( -\frac{\partial \Phi_i}{\partial t}+ mc^2 +\frac{(\nabla
S_i +\frac{q}{c}\vec{A} )^2}{2m}%
-\frac{\hbar^2}{2m}\frac{\Delta\sqrt\rho_i}{\sqrt\rho_i}+\frac{\hbar
q }{2mc}\frac{\chi^\dag\vec\sigma\cdot\vec
H\chi}{\chi^\dag\chi}+U_\chi),
\end{equation}
where
\begin{equation}\label{relativistic corrections phi}
  U_\varphi=
    \frac{\hbar q}{2mc}Re\left(\frac{\varphi^\dag\vec\sigma\vec
  E\chi}{\varphi^\dag\varphi}\right)
\end{equation}
\begin{equation}\label{relativistic corrections chi}
  U_\chi=
    \frac{\hbar q}{2mc}Re\left(\frac{\varphi^\dag\vec\sigma\vec
  E\chi}{\chi^\dag\chi}\right)
\end{equation}

From the definition of the four-velocity:
\begin{equation}\label{auxiliar uno}
  K_\phi=\sum_{i=1,2}\rho_i(-\frac{\partial \Phi_i}{\partial t}-\frac{\partial S_i}{\partial t}-q
  V)
\end{equation}
\begin{equation}\label{auxiliar dos}
K_\chi=\sum_{i=3,4}\rho_i(-\frac{\partial \Phi_i}{\partial
t}-\frac{\partial S_i}{\partial t}+q V)
\end{equation}
where $V$ is the scalar potential of the electromagnetic field.

From these equations and (\ref{energia electron} \& \ref{energia
positron}), we find that the components of $\varphi$ and $\chi$
can be obtained in the low-speed limit as solutions of the system:

\begin{equation}\label{hamilton jacobi electron}
\sum_{i=1,2}%
\rho_i( \frac{\partial S_i}{\partial t}+ mc^2 +\frac{(\nabla
S_i -\frac{q}{c}\vec{A} )^2}{2m}%
-\frac{\hbar^2}{2m}\frac{\Delta\sqrt\rho_i}{\sqrt\rho_i}+ q V+
\frac{\hbar q }{2mc}\frac{\varphi^\dag\vec\sigma\cdot\vec
H\varphi}{\varphi^\dag\varphi}+U_\varphi)=0
\end{equation}
and
\begin{equation}\label{hamilton jacobi positron}
\sum_{i=3,4}%
\rho_i( \frac{\partial S_i}{\partial t}+ mc^2 +\frac{(\nabla
S_i +\frac{q}{c}\vec{A} )^2}{2m}%
-\frac{\hbar^2}{2m}\frac{\Delta\sqrt\rho_i}{\sqrt\rho_i}-qV+\frac{\hbar
q }{2mc}\frac{\chi^\dag\vec\sigma\cdot\vec
H\chi}{\chi^\dag\chi}+U_\chi)=0,
\end{equation}
including (\ref{imag part first} \& \ref{imag part second}).

If we neglect the \emph{zitterbegwegung} term:
\[
\frac{q}{c} \chi^\dag \vec{\sigma}\cdot \vec{E}\varphi\approx 0,
\]
in equations (\ref{imag part first} \& \ref{imag part second}),
considering equations (\ref{madelung transformation}), we can see
that the components of $\varphi$ can be chosen as solutions of a
Schr\"odinger non linear equation([Chapter 3.]\cite{HOLLAND}):
\begin{equation}\label{equation for components of phi}
  i\hbar\frac{\partial \psi}{\partial t}=\frac{(-i\hbar\nabla
  -\frac{q}{c}\vec{A})^2}{2m}\psi+qV\psi+\frac{\hbar q }{2mc}\frac{\varphi^\dag\vec\sigma\cdot\vec
H\varphi}{\varphi^\dag\varphi}\psi+U_\varphi\psi,
\end{equation}
whilst the components of $\chi$ can be chosen as solutions of:
\begin{equation}\label{equation for components of chi}
  -i\hbar\frac{\partial \psi}{\partial t}=\frac{(i\hbar\nabla
  +\frac{q}{c}\vec{A})^2}{2m}\psi-qV\psi+\frac{\hbar q }{2mc}\frac{\chi^\dag\vec\sigma\cdot\vec
H\chi}{\chi^\dag\chi}\psi+U_\chi\psi
\end{equation}
Therefore, at least in the limit where the spinorial part of the
Hamiltonian can be considered small, the components of $\varphi$
and $\chi$ can be chosen as solutions of the non-linear Pauli
equations:
\begin{equation}\label{pauli equation for electron}
 i\hbar\frac{\partial \varphi}{\partial t}=\frac{(-i\hbar\nabla
  -\frac{q}{c}\vec{A})^2}{2m}\varphi+qV\varphi+\frac{\hbar q }{2mc}\vec\sigma\cdot\vec
H\varphi+U_\varphi\varphi,
\end{equation}
and
\begin{equation}\label{pauli equation of chi}
  -i\hbar\frac{\partial \chi}{\partial t}=\frac{(i\hbar\nabla
  +\frac{q}{c}\vec{A})^2}{2m}\chi-qV\chi+\frac{\hbar q }{2mc}\vec\sigma\cdot\vec
H\chi+U_\chi\chi,
\end{equation}
that become linear when $U_\varphi$ and $U_\chi$ are neglected.

Actually, through the Madelung substitutions (\ref{madelung
transformation}), it can be shown that any solution of the system
(\ref{pauli equation for electron}-\ref{pauli equation of chi}) is
a solution of (\ref{hamilton jacobi electron}-\ref{hamilton jacobi
positron}-\ref{imag part first}-\ref{imag part second}), whenever
the \emph{zitterbegwegung} term can be neglected.

In the general, relativistic, case, we recover a picture of
particles moving under the action of the electromagnetic field
alone, although the number of particles is not conserved. Which is
well known to happen, as a matter of fact.

As to the nature of the functions $\Phi_i$, we see that they are
restricted by the condition:
\[
\sum_{i=1,2}\partial^\mu(\rho_i \partial_\mu
\Phi_i)-\sum_{i=3,4}\partial^\mu(\rho_i \partial_\mu \Phi_i)=0,
\]
required by the principle of conservation of electrical charge.
Also, they are solutions of equations (\ref{real part first final}
\& \ref{real part second final}).

\section{Conclusions and Remarks}
If the ideas exposed in this paper are proven to be valid:
\begin{enumerate}
  \item The main assumption of the de Broglie-Bohm theory---that the local
impulse of quantum particles is given by the gradient of the phase
of the wave function---wont be accurate.
  \item However, there will be still room for a classical interpretation of
quantum phenomena, in terms of particles moving along well defined
trajectories, under the action of the electromagnetic field.
\item The number of particles wont be locally conserved.
\item Given that four hidden variables have been introduced---after some considerations on electrodynamics---the
representation of physical systems through wave functions wont be
complete and, therefore, as foreseen by Einstein, Podolsky, and
Rosen, quantum mechanics wont be a complete theory of motion.
\end{enumerate}

\end{document}